\def\be{\begin{equation}}
\def\ee{\end{equation}}
\documentstyle[aps,epsf,prl]{revtex}
\begin{document}
\twocolumn
[
\draft
\title{Coexistence of Haldane gap excitations and long range
antiferromagnetic order in mixed-spin nickelates $R_{2}$BaNiO$_{5}$.}
\author{S. Maslov\thanks{e-mail maslov@cmt3.phy.bnl.gov},
A. Zheludev}
\address{
Physics Department, Brookhaven National Laboratory, Upton, New York
11973}
\date{\today}
\maketitle
\widetext
\begin{abstract}
The spin dynamics of the  $S=1$ Ni-chains in mixed-spin antiferromagnets
Pr$_{2}$BaNiO$_{5}$ and Nd$_{x}$Y$_{2-x}$BaNiO$_{5}$ is described in
terms of a simple Ginzburg-Landau Lagrangian coupled to the sublattice
of rare-earth ions. Within this framework we
obtain a theoretical explanation for the  experimentally observed
coexistence of Haldane gap excitations and long-range magnetic order, as
well as for the increase of the Haldane gap energy below the N\'{e}el
point.
We also predict that the degeneracy of the Haldane  triplet is lifted in
the magnetically ordered phase. The growth of the transverse gap is
shown to follow from the magnon repulsion.
The theoretical results are consistent with the available
experimental data.
\end{abstract}
]
\narrowtext
Much excitement in the field of low-dimensional quantum magnetism was
caused by the theoretical work of Haldane,  who
predicted an energy gap in the excitation spectrum of a
1-dimensional Heisenberg antiferromagnet (1-D HAF) with integer spin
\cite{haldane}. At the
beginning this came as a surprise, since for the half-integer-spin 1-D
HAF the excitation spectrum was known to be gapless. Very soon Haldane's
conjecture was confirmed by many numerical and experimental studies
\cite{haldane2}, and the ``Haldane gap'' phenomenon is by now rather
well understood. A very challenging outstanding problem   now is to
study the {\it quasi}-1-D case, when 3-D magnetic interactions are
sufficiently strong to destroy the Haldane singlet and produce
long-range magnetic order at low temperature, yet some purely
quantum-mechanical effects are preserved, thanks to the dominance of 1-D
interactions. Much experimental and theoretical work in this direction
was done on CsNiCl$_3$\cite{csnicl} and  isostructural integer-spin
compounds \cite{other}. The result was a profound understanding of the
connection that exists between Haldane excitations and ``normal'' spin
waves. In CsNiCl$_3$ for example, the triplet of 1-D Haldane gap modes
is readily observed above the N\'{e}el temperature $T_{N}$. As $T_{N}$ is
approached from above, the gap vanishes at the 3-D magnetic zone-center,
driving a soft-mode transition. In the ordered phase two of the three
modes become conventional gapless spin waves, while the third
``longitudinal'' mode, not accounted for by the linear spin wave theory,
persists as a gapped excitation.

Recently in a series of inelastic neutron scattering studies of quasi-1D
mixed-spin antiferromagnets with the general formula $R_2$BaNiO$_5$
($R=$Pr, Nd or Nd$_x$Y$_{1-x}$)  Zheludev {\it et al.}\cite{zheludev}
demonstrated that
the scenario realized in CsNiCl$_3$ is by no means universal. In
$R_2$BaNiO$_5$ compounds 3-D magnetic ordering produces finite static
moments on both the $S=1$ Ni$^{2+}$ and the $R^{3+}$
magnetic ions \cite{lbanio},
yet the 1-D gap excitations propagating on the Ni-chains are only weakly
affected: they persist in the ordered phase as gapped modes and {\it
coexist} with conventional acoustic spin waves all the way down to
$T=0$. These excitations show no signs of softening at the transition
point at any wave vector. In the entire phase diagram they have a purely
1-D dynamic structure factor that is practically indistinguishable from
that of Haldane excitations in Y$_2$BaNiO$_5$ \cite{ybanio},
a ``clean'' Haldane-gap material, where  only the Ni-sites are
magnetic, and no long-range ordering occurs.
In $R_2$BaNiO$_5$ the only significant (and rather
unexpected) effect of 3-D ordering is that the gap energy {\it
increases} in the ordered phase, the increase being roughly linear with
$(T_{N}-T)$. The survival of 1-D quantum spin  excitations in the
ordered state is indeed remarkable since the transition temperatures are
rather high, and are quite comparable to the gap energy. The purpose of
this letter is to explain the coexistence of long-range order and
Haldane gap excitations in $R_2$BaNiO$_5$ materials.
We propose that, contrary to the case of CsNiCl$_3$,
in these compounds
Ni-chains do not interact between themselves, but instead
each is coupled to the sublattice of magnetic $R$-ions.
We derive the experimentally observed growth
of the gap below $T_N$ using a simple
Ginzburg-Landau Lagrangian \cite{affleck}.
In addition,
we predict that the degeneracy of the Haldane excitations is
partially removed in the ordered phase and notice an
intriguing analogy between this gap splitting in $R_2$BaNiO$_5$ materials
and the famous Higgs mechanism of lifting the mass
degeneracy in particle physics.

Integer spin chains are traditionally described in terms of the
Lagrangian of the Non-Linear Sigma Model (NLSM). In this approach one
approximates the field of spin site-operators $\vec{S}_i$
($\vec{S}_i^2=s(s+1)$), retaining
only the Fourier components near $k_x=0$, and $\pi$ \cite{haldane}. This
is accomplished by a change in real space variables
\begin{equation}
\vec{S}_i = s(-1)^i \vec{\varphi} (i) + \vec{l} (i) \qquad ,
\end{equation}
where both fields $\vec{\varphi} (i)$, and $\vec{l} (i)$ change {\it
slowly} on the scale of one lattice spacing.
According to this definition,
$\vec{\varphi}(i)$ is the unit vector in the
direction of local staggered magnetization, and $\vec{l}(i)$ \ is the
component of local magnetization perpendicular to $\vec{\varphi}(i)$ \
($\vec{\varphi}(i) \cdot \vec{l}(i)=0$).
In the limit of large $s$ three components of $\vec{\varphi}$ commute
with each other, while the commutation rules of $\vec{l}$
remain nontrivial. It can be shown that the correct dynamics
for the new variables follows from a NLSM Lagrangian:
\begin{equation}
{\cal L}= {1 \over 2g} \int dx
\left[{1 \over v} \left( {\partial \vec{\varphi}
\over \partial t} \right)^2- v \left( {\partial \vec{\varphi}
\over \partial x} \right)^2 \right] \qquad ,
\label{nlsm}
\end{equation}
where $\vec{\varphi}$ is subject to
the constraint $\vec{\varphi}^2=1$. Here $g=2/s$ is a
dimensionless coupling constant, measuring the strength of quantum
fluctuations, and $v=2Js$ is the spin wave velocity. The magnetic moment
$\vec{l}$ is given by $\vec{l}={1/gv} \ \vec{\varphi} \times \partial
\vec{\varphi}/\partial t$. This mapping of the 1-D HAF
to the NLSM (\ref{nlsm}) becomes exact for integer $s
\to \infty$, but it gives a meaningful and qualitatively correct
approximation even for $s=1$. From this description it can be derived
that in 1+1 dimensions any strength of quantum fluctuations $g$
is sufficient to destroy long range correlations between spins.
A finite correlation length $\xi$ is always accompanied
by a gap $\Delta = v/\xi$ in the spin excitation spectrum.
For half-integer 1-D HAF the topological term not included in the
Lagrangian
(\ref{nlsm}) prevents the appearance of the finite
correlation length, and spin correlations remain scale-free.
But for integer $s$ this gap, commonly known as the Haldane gap, is
present. It has a small $g$ (large $s$)
asymptotics of $\Delta \sim \exp(-2\pi/g)= \exp(-\pi s)$.
For the  $s=1$ HAF the Haldane gap was found numerically
to be $\Delta=0.41J$ \cite{golinelli}.

The non-linear sigma model is not very convenient for practical
calculations. To simplify things further we follow
the approach of Affleck \cite{affleck},
and replace the Lagrangian of NLSM with that
of the quantum Ginzburg-Landau model
\begin{equation}
{\cal L}= \int dx \left[{1 \over 2v} \left( {\partial \vec{\phi}
\over \partial t} \right)^2- {v \over 2} \left( {\partial \vec{\phi}
\over \partial x} \right)^2 - {\Delta ^2 \over 2v}\vec{\phi} ^2 - \lambda
|\vec{\phi}|^4 \right], 
\label{p4}
\end{equation}
where for convenience we have simultaneously changed variables
to $\vec{\phi}=1/\sqrt{g}\ \vec{\varphi}=\sqrt{s/2}\ \vec{\varphi}$.
The new Lagrangian follows from (\ref{nlsm})
if the constraint $\vec{\varphi}^2=1$ is relaxed, making the vector
of staggered magnetization soft, and phenomenological quadratic and
quartic terms are introduced. The coefficient in front of the
quadratic term is selected to reproduce the correct value of the gap
(in NLSM this gap is generated dynamically),
while the only condition imposed on the quartic term is that
$\lambda>0$, ensuring the overall stability for large $|\vec{\phi}|$.
The physical meaning of Eq. (\ref{p4}) is rather transparent. It
describes the propagation of a triplet of magnons (excitations of
local staggered magnetization) with a given spin wave velocity $v$, and
gap $\Delta$. The quartic term describes the repulsive
magnon-magnon interaction.

The Lagrangian (\ref{p4}) was successfully used to qualitatively explain
the behavior of coupled 1D spin chains in CsNiCl$_3$ \cite{affleck}.
In this case the $i$-th chain is
described by Eq. (\ref{p4}) with $\vec{\phi} \to
\vec{\phi}_i$, while the interchain coupling is given by
$-2s J' \vec{\phi}_i \cdot \vec{\phi}_{i+1}$. On the mean-field level
the behavior of the excitations is the following: the gap is reduced by the
interchain coupling to  $\Delta_{eff}^2=\Delta^2 (T)-4vZsJ'$, where $Z$ is
the number of chains coupled to a given chain. The O(3) symmetry is
spontaneously broken for $\Delta^2 (T)<4ZsvJ'$. In this case
$|\vec{\phi|}$ has a nonzero expectation value of
$\sqrt{(4vZsJ'-\Delta^2 (T))/4 v \lambda}$.
The gap naturally grows with temperature and eventually
wins over the interchain exchange, restoring the O(3) symmetry. The
N\'{e}el temperature is determined by the condition $\Delta^2 (T_N)=
4vZsJ'$. It is important to understand what happens to the triplet of
magnons below $T_N$. As $T_N$ is approached from above, the gap in all
three excitations at the positions of future magnetic Bragg peaks
decreases and hits zero at $T_N$.
Below $T_N$ two excitations corresponding to the two Goldstone modes of
the N\'{e}el order parameter remain gapless. At the same time the
longitudinal branch re-acquires the gap along with a finite
lifetime due to decay into two gapless modes.

The ``traditional'' scenario \cite{affleck} described above is in
agreement with experimental findings in CsNiCl$_3$. However, for
$R_{2}$BaNiO$_{5}$ we can expect a different picture. The structure of
these materials is such that there  are no strong direct superexchange
paths between the Ni chains, which explains why in isostructural
Y$_2$BaNiO$_5$ no magnetic ordering is observed \cite{ybanio}. On the
other hand, magnetic  rare earth ions substituted for Y are coupled
between themselves. To a good approximation one can assume that these
sites form a separate sublattice which orders at rather high
temperatures ($T_N=48$K for Nd$_2$BaNiO$_5$). The coupling $J'$ between
the Ni  and $R$ sublattices  is weak, yet finite. The finite ordered
moments on the Ni-sites  may be interpreted as the singlet-ground-state
Ni-chains being {\it polarized} by a staggered exchange field from the
magnetically ordered sublattice of $R$-ions. To describe this effect an
``external field'' term should be added to the Ginzburg-Landau
Lagrangian~(\ref{p4}). It is given by $J'
\vec{n} \cdot \sqrt{2s} \vec{\phi}=\vec{H}_s \cdot
\vec{\phi}$, where $\vec{n}$ is the staggered moment of the
$R$ sublattice.
Such a term explicitly breaks the $O(3)$ symmetry and
causes $\vec{\phi}$ to acquire a  nonzero expectation value along
$\vec{H}_s$. As a consequence, the degeneracy of the triplet of Haldane gap
excitations is partially lifted. The energy gap in the magnon, polarized
parallel to $\vec{H}_s$, is different from that in the two transversal
branches. Both gaps can be derived from ${\cal P}$ -- the potential
energy density of the Lagrangian:
\begin{eqnarray}
{\cal P}= {1 \over 2v} \left( {\partial \vec{\phi}
\over \partial x} \right)^2+{\Delta ^2 \over 2v}\vec{\phi} ^2 + \lambda
|\vec{\phi}|^4 - \vec{H}_s \vec{\phi} \qquad .
\label{pot}
\end{eqnarray}
The usual formula for this is
$\Delta_{\alpha}^2/v=\partial^2 {\cal P}/ \partial \phi_{\alpha}^2$.
It gives
\begin{eqnarray}
\Delta_{||}^2=\Delta_0^2+12 v \lambda \phi_0^2 \quad ; \nonumber \\
\Delta_{\perp}^2=\Delta_0^2+4 v \lambda \phi_0^2 \qquad .
\end{eqnarray}
Here $\phi_0$ is the expectation value of $|\vec{\phi}|$ determined by
minimization of (\ref{pot}) through
\begin{equation}
H_s=\Delta_0^2/v  \ \phi_0 + 4 \lambda \phi_0^3 \qquad ,
\label{op}
\end{equation}
or, approximately, $\phi_0 \simeq vH_s/\Delta_0^2$. Note that
both energy gaps are increased compared to their value in the absence of
the staggered field. This observation relies on the fact that $\lambda
>0$, i.e. a repulsive magnon-magnon interaction. It is precisely the
positive $\lambda$, which guaranties the stability of Eq. (\ref{op})
and of the Lagrangian (\ref{p4}) itself, and, therefore, should
be satisfied.

As we mentioned in the introduction, the Lagrangian of NLSM
is traditionally used to describe the behavior of quantum spin chains.
The effect of the staggered magnetic field in NLSM
, coupled to the staggered magnetization via
the term $\vec{n} \cdot \vec{h}/g$,
was studied by Nelson and Pelcovits \cite{np} using renormalization
group methods. Their RG flow indicates that indeed the gap
increases (the correlation length decreases) in strong enough
fields. The one loop corrections calculated by them do not give a
definite answer about the behavior of the correlation length in weak
fields. Using our results for the Ginzburg-Landau Lagrangian
we can conclude that the correlation length decreases in this
case as well.

Our predictions for the energy gap can be qualitatively understood by
working with the original Heisenberg Hamiltonian and treating the
staggered field as a weak perturbation. Let us denote the Haldane ground
state  by $|G\rangle$ and label the lowest-energy triplet excitations at
$q=\pi$ as $|E,+1\rangle$, $|E,0\rangle$ and $|E,-1\rangle$, according
to their value of $S_{z}$. Let us now consider the relevant matrix
elements of the staggered field operator
$\hat{H}_{s}=H_{s}\sum_{j}(-1)^{j}S^{z}_{j}$. These are non-zero only
between states with momentums differing by exactly $\pi$.  In
particular, $\langle G|\hat{H}_{s}|G\rangle$, as well as  the matrix
elements between any two of the $|E,+1\rangle$, $|E,0\rangle$ or
$|E,-1\rangle$ states are strictly zero.  As a result, there are no
first-order corrections to the spin gap at $q=\pi$. To calculate the
2-nd order correction $\delta E_G$ to the ground state energy we can use
the well-known fact that the triplet of Haldane modes pretty much
exhausts the spectral weight at $q=\pi$. Thus, only the mixing of
$|G\rangle$ with single-particle excitations needs to be considered.
Since $\hat{H}_{s}$ conserves the z-component of total spin, its only
non-zero matrix element is the one between $|G\rangle$ and $|E,0
\rangle$. The correction to the ground state
energy is thus given by $\delta E_G=-\frac{|\langle E, 0|\hat{H}_{s}|G
\rangle|^2}{\Delta}\equiv -\epsilon$. We now proceed to calculate the
energy corrections
for the three excited states at $q=\pi$. Two effects need to be
considered: i) their mixing with the ground state and ii) their mixing
with 2-particle excited states. Let us first assume that there is no
repulsion between magnons, and that doubly-excited states at $q=0$,
which we label as $|2E,+1,0\rangle$, $|2E,-1,0\rangle$, and $|2E, 0,
0\rangle$, are produced  by adding two non-interacting $q=\pi$
excitations to the system. Second order corrections to the energies of
singly-excited states are then given by: $\delta E_{E,0}=\frac{|\langle
G|\hat{H}_{s}|E, 0 \rangle|^2}{\Delta}-
\frac{|\langle 2E, 0, 0|\hat{H}_{s}|E, 0
\rangle|^2}{\Delta}=\epsilon-\epsilon=0$ and
$\delta E_{E,+1}=-\frac{|\langle 2E, +1, 0|\hat{H}_{s} |E, +1
\rangle|^2}{\Delta}=-\epsilon=\delta E_{E,-1}$.  The correction to the gap
energies then become {\it zero} for the transverse gap $\Delta_{\bot}$,
and $+\epsilon\propto H_{s}^2$  for the longitudinal gap $\Delta_{\|}$.
Magnon repulsion effectively reduces the negative second-order
corrections to the single-particle excitation energies that are due to
mixing  with double-excited states. Even for the transverse modes magnon
repulsion will thus lead to an {\it increase} of the gap energy. Both
$\Delta_{\bot}$ and $\Delta_{\|}$ will increase quadratically with
$H_{s}$, though for $\Delta_{\|}$ the effect will be more pronounced.
This is totally consistent with results derived above, where the
positive parameter $\lambda$ represents repulsive interactions between
magnons.

An increase of the energy gap below the transition point  is exactly
what was observed in  neutron scattering experiments on
Nd$_2$BaNiO$_5$, Nd$_{\rm x}$Y$_{2-\rm x}$BaNiO$_5$, and
Pr$_2$BaNiO$_5$ \cite{zheludev}. The data
were taken on powder samples or single crystals of small size, which
obviously made resolving the  mode splitting below $T_N$ impossible. In
the powder measurements at least, the observed  gap always corresponds
to the lowest-energy mode. In Fig. 1 we show the combined data from all
experiments,  plotting the increase of the square of gap energy
(relative to that
measured in Y$_{2}$BaNiO$_{5}$ \cite{gap_ybanio} at the same
temperature) against the square of
staggered magnetization of the Ni sublattice.  The data are consistent
with the predicted linear dependence. Moreover, data points for
different  compounds seem to collapse onto a single curve,
 even though the N\'{e}el temperatures vary considerably.
The  mechanism is therefore insensitive to the details of
exchange in the $R$-sublattice, and between Ni and $R$ sublattices.

In a much obscured form the effect of staggered field has been
previously observed in the well-known Haldane-gap material NENP
\cite{chiba}. In this  compound applying an external {\it uniform} field
$H$ produces a weak effective staggered component due to some special
structural features. On top of a very pronounced linear splitting of the
Haldane triplet by the uniform field  for each mode one observes a
slight energy shift that is {\it positive} and quadratic with $H$. This
shift was attributed to the weak staggered field component that for NENP
is inseparable from the uniform one \cite{sakai}. Numerical calculations
have confirmed this conclusion \cite{mitra}. Note that unlike in NENP,
in $R_{2}$BaNiO$_{5}$ systems we are able to see the effect of staggered
field much more clearly, since the dominant uniform field component is
absent.

It is interesting to mention that the effect of gap increase and
splitting due to interaction with another sublattice has a close
analogue in the field of particle physics. It is very similar to the
Higgs mechanism by which the mass degeneracy in  a multiplet of
elementary particles is lifted. As in our case, the splitting of
masses
(energy gaps  relative to a vacuum of particles) is a result of spontaneous
symmetry breaking, produced  by the interaction with Higgs particles. The
three Ni chain Haldane excitations of different polarizations play the
role of a multiplet of particles of initially equal masses (quarks for
example), while the conventional acoustic spin waves, referred to as
mixed $R$-Ni excitations by Zheludev {\it et al.},  play the role of the
elusive Higgs particle, which is massless in our case.
This analogy should not be taken too seriously
due to the differences in Hamiltonians: it is the SU(2) symmetry which
is spontaneously broken by the Higgs mechanism, and the term describing the
coupling between matter and Higgs particles is more complicated than a
simple $-\vec{H}_s \vec{\phi}$ term in our case.
Nevertheless, the essence of the effect is the
same: spontaneous symmetry breaking caused by interaction with some
external field lifts the degeneracy of masses (gaps).

In conclusion, we have provided a simple theoretical explanation for the
coexistence of long range magnetic order and Haldane gap excitations
in $R_2$BaNiO$_{5}$ systems. The Ni-chain gap  modes are described in
terms of a simple Ginzburg-Landau Lagrangian. The increase of the gap
below the N\'{e}el temperature, previously observed experimentally in
several compounds, immediately follows from this approach. The
qualitatively new prediction  is that the degeneracy of the Haldane
triplet is lifted below the transition point. The magnon polarized along
the vector of staggered magnetization has a larger gap than the two
others magnon branches. Experimental tests of this prediction, including
inelastic neutron scattering experiments on large high-quality single
samples, are planned for the near future.

We thank Victor Emery and Igor Zaliznyak for helpful discussions,
and V.N. Muthukumar for bringing Ref.
\cite{sakai,mitra} to our attention.
This work  was carried out under Contract No. DE-AC02-76CH00016,
Division of Material Science, U.S.\ Department of Energy.

\begin{figure}
\caption{Experimentally measured increase of the
square of the gap energy in $R_{2}$BaNiO$_{5}$
samples relative to that in Y$_{2}$BaNiO$_{5}$
at the same temperature, plotted against the
square of staggered magnetization of Ni sites.}
\end{figure}

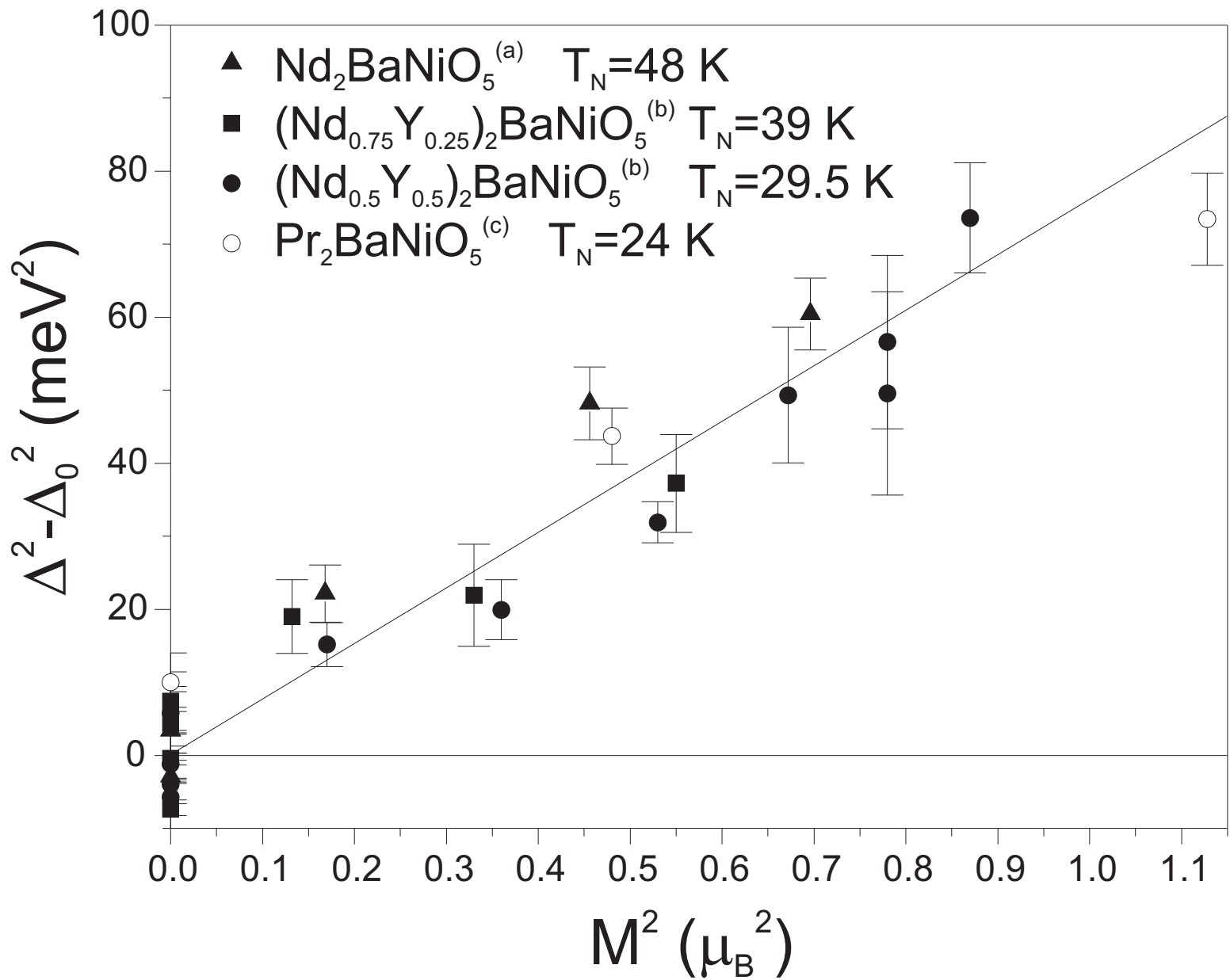
\end{document}